\documentclass[9pt,twocolumn,twoside]{osajnl}
\journal{ol} 
\setboolean{shortarticle}{true}
\usepackage{siunitx}
\usepackage{pdfpages}

\title{Spatio-temporal observation of higher-order modulation instability in a recirculating fiber loop}

\author[1,*]{François Copie}
\author[1]{Pierre Suret}
\author[1]{Stephane Randoux}

\affil[1]{Univ. Lille, CNRS, UMR 8523 - PhLAM - Physique des Lasers Atomes et Molécules, F-59000 Lille, France}

\affil[*]{Corresponding author: francois.copie@univ-lille.fr}

\begin{abstract}
	We experimentaly investigate higher-order seeded modulation instability in an optical fiber experiment. The recirculating loop configuration with round-trip losses compensation enables the observation in single-shot of the spatio-temporal evolution of an initially modulated continuous field revealing intricate yet deterministic dynamics. By tuning the modulation period, a continuous transition between perfectly coherent and purely noise-driven dynamics is observed that we characterise by means of a statistical study. 
\end{abstract}

\begin{document}

	\maketitle

\begin{figure}[htbp]
	\centering
	\includegraphics[width=\linewidth]{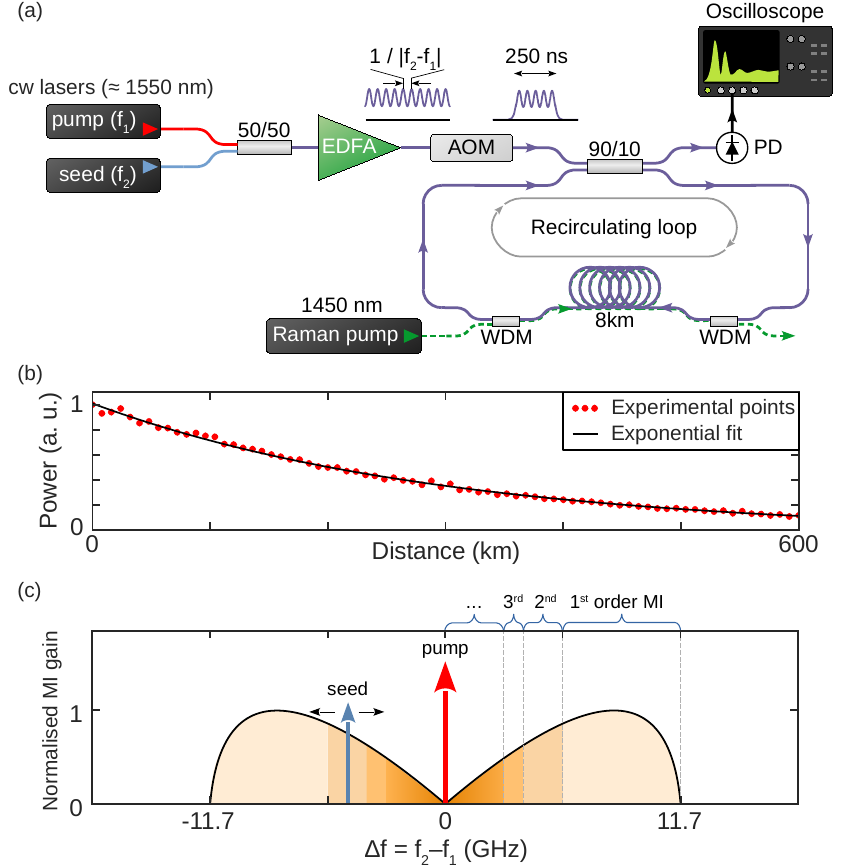}
	\caption{(a) Schematic of the experimental setup. (b) Evolution of the power decay measured in an experimental run. (c) Small signal MI gain computed from experimental parameters. $P_p \sim \SI{24}{mW}$, $\gamma = \SI{1.23}{/W/km}$, $\beta_2(\SI{1550}{nm}) = \SI{-22}{ps^2/km}$, $\alpha_{\text{eff}} = \SI{3.36 e-6}{m^{-1}}$. EDFA: Erbium doped fiber amplifier, AOM: Acousto-optic modulator, PD: Photodetector, WDM: Wavelength division multiplexer.}
	\label{fig:1}
\end{figure}

Modulation Instability (MI) is one of the most ubiquitous phenomena of nonlinear dynamics owing to its appearance in very diverse fields including hydrodynamics, Bose-Einstein condensate physics or optics to name a few  \cite{zakharov_modulation_2009}. It occurs when a pump field interacts with a weaker seed field and manifests itself, at early stage, by the exponential growth of the seed provided that its frequency falls below a certain cutoff which defines the so-called MI gain band. In optical fiber systems, one of the most common scenario encountered is the noise-induced or spontaneous MI in which a single continuous wave (cw) pump field perturbed by spectral noise breaks up into a partially stochastic train of short pulses before experiencing a complex spatio-temporal dynamics \cite{tai_observation_1986, agrawal_nonlinear_2013}. Another common configuration is the coherently seeded MI which is triggered by a weak modulation of a cw pump laser and that leads to the formation of well defined pulse trains \cite{hasegawa_generation_1984}. At longer propagation distance, the dynamics triggered by seeded MI can feature quasi-periodic spatio-temporal evolutions of the wavefield that have been linked to the Fermi-Pasta-Ulam-Tsingou (FPUT) recurrences \cite{trillo_dynamics_1991}. More specifically, such a parallel is drawn only for `1\textsuperscript{st}-order MI' which is a subset of the seeded MI in which the modulation frequency of the cw field lies in the outer half of the MI gain band (See Fig. \ref{fig:1}(c)). Conversely, so-called higher-order MI can arise from a single modulation whose frequency falls below half of the cutoff frequency and results in considerably different spatio-temporal dynamics of the wavefield \cite{wabnitz_efficient_2010, erkintalo_higher-order_2011, hammani_peregrine_2011, kimmoun_nonconservative_2017}. Indeed, in that case one or several harmonics of the modulation frequency also fall in the MI gain band generated by the pump which leads to intricate multiple wave mixing. The early stage of the dynamics features a generic splitting cascade that has been succesfully described analytically as resulting from the nonlinear superposition of elementary breathers \cite{erkintalo_higher-order_2011}. Yet, the long term dynamics of higher-order MI remains largely unexplored.

Until relatively recently, MI has been the subject of numerous experimental observations in fiber based systems mainly limited to measurements of averaged quantities such as conventional optical spectra or autocorrelation traces for instance due to the intrinsic rapid timescale associated to the process \cite{tai_observation_1986, hammani_spectral_2011}. In the last 10 years though, the development of single-shot detection systems and distributed measurement methods based in particular on dispersive Fourier transformation (DFT), time-lens systems and time-domain reflectometry has attracted particular interest and enabled, among other things, the precise characterisation of MI-driven ultrafast dynamics \cite{solli_fluctuations_2012, suret_single-shot_2016, narhi_real-time_2016, tikan_single-shot_2018, lebel_single-shot_2021, mussot_fibre_2018, naveau_heterodyne_2021}. In the past few years, the recirculating optical fiber loop configuration (historically developed to study long-haul transmissions of solitonic pulses \cite{desurvire_raman_1985, mollenauer_demonstration_1988}) has also proven to be a powerful experimental platform to observe and investigate complex spatio-temporal dynamics induced by MI \cite{kraych_nonlinear_2019, kraych_statistical_2019, goossens_experimental_2019}.

In this work, we report original observations of higher-order MI realised in a recirculating optical fiber loop. We recorded the spatio-temporal evolution of long square pulses sinusoidally modulated that propagate over hundreds of kilometers, exhibiting a variety of remarkable scenarios. It generally consists in a pulse splitting sequence followed by a complex coherent pulse recombination dynamics. When changing the frequency of the initial modulation, we were able to reveal a controlled interplay between higher-order seeded MI and the noise-induced MI thanks to our single-shot recording technique.

The evolution of optical wavefields propagating in our recirculating fiber loop is well described by the following 1D nonlinear Schrödinger (NLS) equation including effective linear dissipation:
	\begin{equation}
	i \frac{\partial \Psi}{\partial z} = \frac{\beta_2}{2}\frac{\partial^2 \Psi}{\partial t^2} - \gamma |\Psi|^2 \Psi - i \frac{\alpha_{\text{eff}}}{2} \Psi
	\label{eq:NLSE}
	\end{equation}
	
$\Psi(z, t)$ is the envelop of the field, $\beta_2$ and $\gamma$ are the group velocity dispersion (GVD) and nonlinear coefficients at \SI{1550}{nm} respectively, $\alpha_{\text{eff}}$ is the effective power loss coefficient, $z$ is the dimensional propagation distance and $t$ the time in the frame travelling at the group velocity of the pump field. The fiber used in our work exhibits anomalous GVD at the pump wavelength ($\beta_2 < 0$) which is refered to as the focusing regime \cite{copie_physics_2020}. For the parameters of our experiments, the MI gain curve given by $g(\Delta f) = |\beta_2 \Delta f|\sqrt{f_c^2 - \Delta f^2}$, where $f_c = \sqrt{(4 \gamma P_p)/|\beta_2|}/(2\pi)$ is the MI cutoff frequency, is plotted in Fig.\,\ref{fig:1}(c) \cite{agrawal_nonlinear_2013}. $\Delta f$ is the pump-seed frequency detuning which is varied in our experiments by changing the frequency of a seed laser around that of a pump laser of power $P_p$. The power decay during propagation (longitudinal decrease of $|\Psi|^2$) influences the MI gain curve which varies along propagation. Note that the expression of $g(\Delta f)$ is derived analytically in the case of an initially infinitely small modulation, i.e. for very small seed power. In the presented results, the power of the seed field is \SI{9}{dB} lower than that of the pump field.
	
Our experimental setup is schematically depicted in Fig.\,\ref{fig:1}(a). Two cw lasers respectively called pump and seed are mixed via a 50/50 coupler before being amplified by an Erbium doped fiber amplifier (EDFA). This gives a modulated signal at a beating frequency given by the detuning between the two lasers. This field is then modulated by an acousto-optic modulator (AOM) to generate long pulses (\SI{250}{ns}) which mitigate Brillouin scattering. The pulses are then launched into an \SI{8}{km}-long recirculating loop made of standard single mode fiber. At each round-trip, 10\% of the circulating power is extracted and directed towards a photodetector (PD) coupled to a sampling oscilloscope leading to an overall \SI{32}{GHz} detection  bandwidth. The losses accumulated over one circulation in the fiber loop are partially compensated thanks to a counter-propagating Raman pump coupled in and out of the loop via wavelength division multiplexers (WDMs). This reduces the effective power decay rate of the circulating field to \SI{0.015}{dB/km} or equivalently $\alpha_{\text{eff}} \sim \SI{3.36 e-6}{m^{-1}}$ (see Fig.\,\ref{fig:1}(b)). Data recorded with the oscilloscope consist in a succession of sequences (one per round-trip) that are numerically-processed to construct single-shot real-time spatio-temporal diagrams of the wavefield dynamics. By tuning the frequency of the seed around that of the pump, the system is set in an initial state where different orders of MI can be observed.

\begin{figure*}[t]
	\centering
	\includegraphics[width=\linewidth]{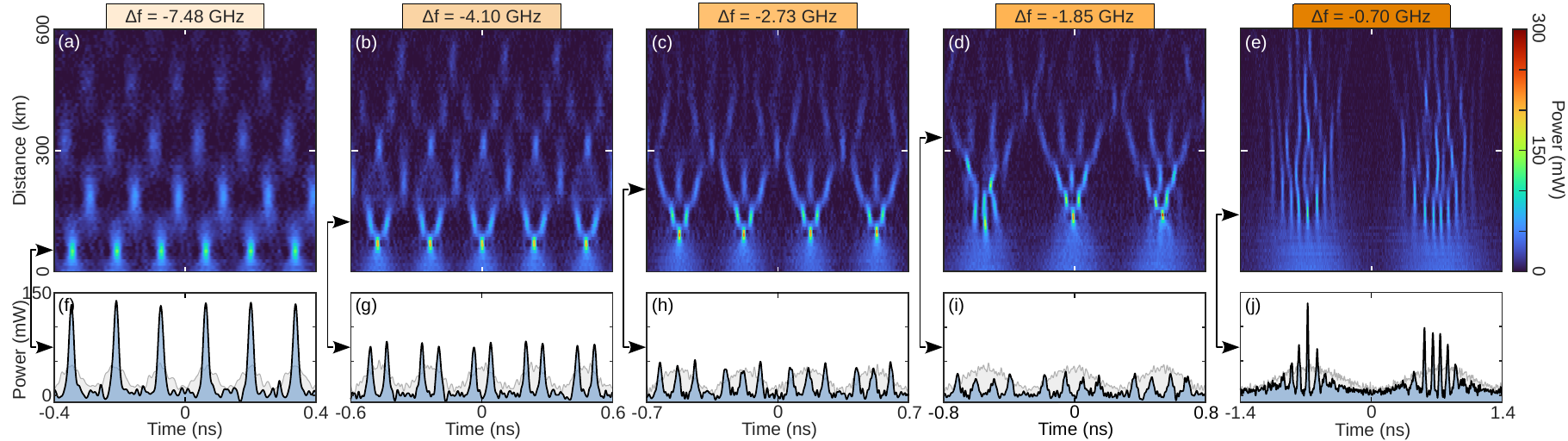}
	\caption{(a-e) Experimental recordings of the spatio-temporal evolution of initially modulated wavefields for different frequency detunings. (f-j) Temporal traces extracted from (a-e) at the propagation distances indicated by the arrows. The traces in light gray on the background are taken after one circulation (\SI{8}{km}). Experiments reported in (a-e) are associated to ratios $f_c / \Delta f = 1.6,~ 2.8,~ 4.3,~ 6.3$ and $16.7$ respectively, $f_c$ being estimated at the initial condition (for $P_p \sim \SI{24}{mW}$).}
	\label{fig:2}
\end{figure*}

We have recorded several hundreds of spatio-temporal diagrams covering the full MI bandwidth by tuning the seed frequency with a step of $\sim \SI{125}{MHz}$. Figure \ref{fig:2} shows excerpts of our experimental results obtained for five values of frequency detuning ranging from -7.48 to \SI{-0.7}{GHz}. The corresponding real-time spatio-temporal dynamics of light intensity are presented as 2D color plots in Fig.\,\ref{fig:2}(a-e).  In each case, the temporal window showed in the plots (horizontal axis) is restricted to a fraction ($\sim \SI{1}{ns}$) of the actual recorded length ($>\SI{250}{ns}$) for clarity. For $\Delta f = \SI{-7.48}{GHz}$ (Fig.\ref{fig:2}(a)), the ratio $f_c / \Delta f$ calculated from the initial condition is $\sim 1.6$ which means that only 1\textsuperscript{st} order MI can arise. The observed spatio-temporal dynamics is reminiscent of the Fermi-Pasta-Ulam-Tsingou recurrences that are usually observed within this parameter range \cite{goossens_experimental_2019, vanderhaegen_observation_2020-2}. Self-compression of the modulated field leads to the generation of a high contrast train of pulses (see Fig.\,\ref{fig:2}(f)) followed by a return close to the initial state. This dynamics is repeated quasi-periodically along propagation. Note that in our case, the spectrum of the initial condition is asymetric, thus the quasi recurrences are neither in-phase, nor out-of-phase \cite{trillo_dynamics_1991} but rather ``continuously'' drift towards long times.

For $\Delta f = \SI{-4.10}{GHz} ~\text{and}~ \SI{-2.73}{GHz}$ (Fig.\,\ref{fig:2}(b, c) respectively), the initial harmonic modulation self-compresses towards a coherent pulse train but each of the pulses later experiences a coherent splitting dynamics which is the signature of higher order MI \cite{erkintalo_higher-order_2011, hammani_peregrine_2011}. When $\Delta f = \SI{-4.10}{GHz}$, a clear split in two is observed after \SI{120}{km} of propagation (Fig.\,\ref{fig:2}(g)) which precedes a complex yet still remarkably coherent dynamical evolution of the pulse train. For $\Delta f = \SI{-2.73}{GHz}$, the splitting cascade continues further as the initial modulation turns into regular triplets of pulses after \SI{224}{km} (Fig.\,\ref{fig:2}(h)). Importantly, we note that the initial ratios $f_c / \Delta f$ for these 2 recordings are 2.8 and 4.3 which means that at least 2\textsuperscript{nd} and 4\textsuperscript{th} order MI should occur respectively in the conservative case. In practice, we indeed observe a dynamics compatible with 2\textsuperscript{nd} order MI for $\Delta f = \SI{-4.10}{GHz}$ but only 3\textsuperscript{rd} order MI for $\Delta f = \SI{-2.73}{GHz}$ as a consequence of dissipation. Indeed, the exponential decay of the power reduces the cutoff frequency of the MI gain which decreases the ratio $f_c / \Delta f$ during propagation such that the initial value of this ratio necessarily surestimates the actual order of MI that can be observed when there is dissipation. It is important to point out that evolutions very similar to the regular splitting sequence observed in our experiments appear during the propagation of higher-order solitons and more generally in the evolution of any arbitrary shaped pulse of sufficiently large peak power \cite{agrawal_nonlinear_2013, hammani_peregrine_2011, tikan_universality_2017, tikan_effect_2020, bertola_universality_2013}. This suggests that the spatio-temporal dynamics induced by higher-order MI at early stage is almost exclusively dictated by the local gradient (hump) of the initial optical power, i.e. there is no coupling between adjacent modulation periods. 

When the frequency detuning is further decreased, the coherence of the spatio-temporal dynamics is partially lost as shown in Fig.\,\ref{fig:2}(d) due to an emergent competion with noise-induced MI. Indeed, we observed that each hump of the initial condition exhibits a different spatio-temporal dynamics and often does not experience compression to a single pulse followed by a regular splitting cascade. As an example, the hump on the left part of Fig.\,\ref{fig:2}(d) shows a complex dynamics that features a degree of stochasticity higher than its neighbours. Noticeably though, the field after \SI{328}{km} of propagation forms almost perfectly regular quartets of pulses (Fig.\,\ref{fig:2}(i)). Numerical simulations allows to  impute this to the effective dissipation of the system. For a frequency detuning as low as $\SI{-0.70}{GHz}$, the spatio-temporal dynamics is completely driven by spontaneous MI as depicted in Fig.\,\ref{fig:2}(e). Indeed, the frequency of the initial modulation is nearly 10 times lower than the frequency of maximum MI gain which means that spectral noise close to the maximum gain experiences a growth rate significantly larger than that of the seed field. Each hump of the initial modulation evolves as a slowly modulated wavefield that breaks up into a train of almost regular pulses (see Fig.\,\ref{fig:2}(j)) that later evolves stochasticaly. Interestingly, the maximum contrast of the pulse train is obtained earlier at the times where the initial modulation is maximum. A related behaviour has been described numerically and analytically in Ref. \cite{erkintalo_akhmediev_2011} using the Akhmediev breather formalism but the present observation shows that it is a more general phenomenon because it is also relevant to the case of spontaneous MI. Note that for $\Delta f > 0$, we get symetrical observations to those reported in Fig.\,\ref{fig:2}. Also, we have performed similar experiments with greater pump-to-seed power ratio which resulted in qualitatively identical observations. 

\begin{figure}[htbp]
	\centering
	\includegraphics[width=\linewidth]{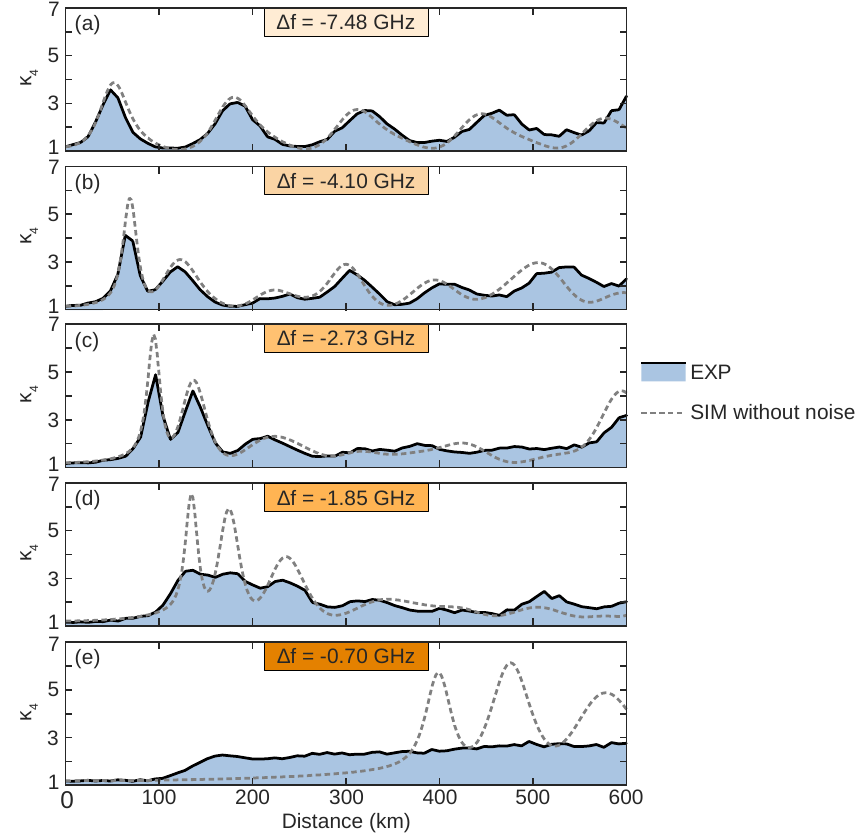}
	\caption{Evolution of the second order moment $\kappa_4$ calculated from the experimental recordings illlustrated in Fig.\,\ref{fig:2}. Evolution of $\kappa_4$ computed from numerical simulations without inclusion of noise is superimposed as dashed grey lines. $P_p = \SI{24}{mW}$, $P_s = \SI{3}{mW}$, $\alpha_{\text{eff}} = \SI{3.36 e-6}{m^{-1}}$.}
	\label{fig:3}
\end{figure}

The gradual loss of coherence that is observed in our experiments originates from the competition between seeded and spontaneous MI that is mediated, for a fixed level of spectral noise, by the frequency detuning $\Delta f$. Note that it has been shown numerically that a similar interplay could disrupt the FPUT recurrences induced by 1\textsuperscript{st} order MI \cite{wabnitz_instability_2014}. In order to get a quantitative insight into this transition we focus on the evolution of a statistical indicator which is the second-order moment of the power $\kappa_4(z) = \left\langle P(z, t)^2 \right\rangle / \left\langle P(z, t) \right\rangle^2 $, where $\left\langle \cdot \right\rangle$ denotes average over time $t$. Larger values of $\kappa_4(z)$ traduce the presence, at given distance $z$, of an increased number of localised structures having high peak power. In the case of cw spontaneous MI, the value of $\kappa_4(z)$ varies from 1 at initial state to 2 at long propagation distance and features damped oscillations with a first overshoot that pinpoints the statistical position of the first maximum compression of the pulse train generated by MI \cite{kraych_statistical_2019}. Figure \ref{fig:3} shows the longitudinal evolution of $\kappa_4(z)$ calculated for the 5 values of $\Delta f$ presented in Fig.\,\ref{fig:2} as blue shaded areas. For comparison, the evolution of $\kappa_4(z)$ computed from numerical simulations of the NLS equation without inclusion of spectral noise is superimposed as dashed gray lines. The initial field for the simulation is $\Psi(0, t) = [\sqrt{P_p} + \sqrt{P_s}\exp(i 2\pi \Delta f \times t)]$ where $P_p$ and $P_s$ are the initial pump and seed powers respectively. Figure \ref{fig:3} shows that our statistical treatment provides clear identification of the influence of spontaneous MI over the dynamics of seeded MI. In the case of 1\textsuperscript{st} order MI (Fig.\,\ref{fig:3}(a)), the evolution of $\kappa_4$ presents 4 periods traducing the 4 quasi-recurrences observed in Fig.\,\ref{fig:2}(a). In Figs. \ref{fig:3}(b-c) relative to $\Delta f = \SI{-4.10}{GHz}$ and $\SI{-2.73}{GHz}$, the evolution of $\kappa_4$ features a first strong peak (first maximum of compression) followed by a second one (first coherent splitting) and even a weak third one (second coherent splitting) in Fig.\,\ref{fig:3}(c). This is in adequation with our identification of 2\textsuperscript{nd} and 3\textsuperscript{rd} order MI dynamics in the spatio-temporal diagrams of Fig.\,\ref{fig:2}(b-c). Note that the apparent ``fluctuations'' of $\kappa_4$ observed in Fig.\,\ref{fig:3}(b) after \SI{250}{km} actually still traduce a coherent dynamics (see Fig.\,\ref{fig:2}(b)) but altered by linear dissipation. The very good agreement between experiments and numerical simuations without noise in these first 3 cases indicates that the dynamics is almost uniquely governed by seeded MI. Consistently with the results presented previously, spontaneous MI dramaticaly alters the coherent dynamics of the wavefield when $|\Delta f| < \SI{2}{GHz}$. This translates in the longitudinal evolution of $\kappa_4$ shown in Fig.\,\ref{fig:3}(d) by the fact that the large oscillations in the first \SI{300}{km} of propagation observed in the numerical simulation without noise are almost completely smoothen out in experiments. For even smaller frequency detuning (Fig.\,\ref{fig:3}(e)), coherent oscillations of $\kappa_4$ vanish and we observe a weak overshoot at \SI{160}{km} that is reminiscent of spontaneous MI. A complete picture of the evolution of $\kappa_4(z)$ when scanning $\Delta f$ is presented as a 2D-color plot in the Supplemental document along with corresponding numerical simulations to further illustrate the progressive competition between seeded and spontaneous MI. 

\begin{figure}[htbp]
		\centering
		\includegraphics[width=\linewidth]{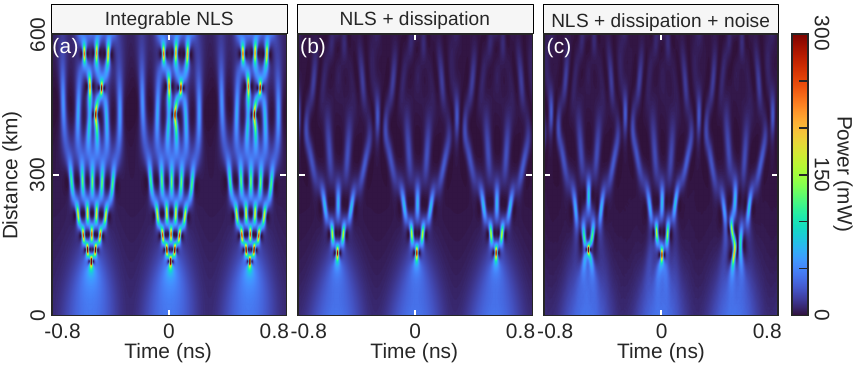}
		\caption{Numerical simulations of Eq. (\ref{eq:NLSE}) of the experiment presented in Fig.\,\ref{fig:2}(d). (a) Without dissipation nor initial noise, (b) with linear dissipation, (c) with linear dissipation and inclusion of initial noise (realistic simulation).}
		\label{fig:4}
	\end{figure}

Finally, we analyse through numerical simulations the influence of realistic dissipation and noise on the spatio-temporal evolution of the wavefield when higher-order MI is involved. To do so, we focus on the case $\Delta f = \SI{-1.85}{GHz}$ illustrated in Figs. \ref{fig:2}(d) and \ref{fig:3}(d) as it particularly highlights the interplay between seeded and spontaneous MI. Figure \ref{fig:4}(a) shows the result of a simulation of the integrable NLS equation (i.e. with $\alpha_{\text{eff}} = 0$) and no added noise. The longitudinal evolution up to $\SI{300}{km}$ shows a regular splitting cascade up to 5\textsuperscript{th}-order before the spatio-temporal dynamics becomes more complex while preserving its temporal coherence. Note that the observed temporal asymmetry is solely due to the single-sideband excitation. The addition of linear dissipation to the simulation changes quantitatively the observed result (see Fig.\,\ref{fig:4}(b)) since the coherent splitting cascade is stopped at 4\textsuperscript{th}-order. Additional inclusion of realistic noise to the initial condition results in the spatio-temporal dynamics illustrated in Fig.\,\ref{fig:4}(c) which agrees very well with our experimental observation of Fig.\,\ref{fig:2}(d). The spectral noise added to the initial state triggers spontaneous MI which partially disrupt the coherence of the higher-order MI.

In conclusion, we have used a recirculating optical fiber loop system to observe in real-time the spatio-temporal evolution of quasi-continuous wavefields experiencing seeded higher-order modulation instability in a recirculating optical fiber loop system. Fine scanning of the frequency detuning between a pump and seed laser enabled clear observation of the occurence of different orders of modulation instability which result in a characteristic coherent pulse splitting cascade. At low frequency detuning, our experiments reveal the coexistence of seeded and spontaneous MI, of which we analyse the interplay using statistical tools.\\

	\noindent \textbf{Funding.} Agence Nationale de la Recherche (StormWave ANR-21-CE30-0009 and SOGOOD ANR-21-CE30-0061 projects, LABEX CEMPI ANR-11-LABX-0007, I-SITE ULNE ANR-16-IDEX-0004); Ministry of Higher Education and Research, Hauts de France council and ERDF through the CPER P4S.\\
	
	\noindent \textbf{Disclosures.} The authors declare no conflicts of interest.\\\vspace{-1em}
	
	\bibliography{Article_HOMI}
	\bibliographyfullrefs{Article_HOMI}



\section*{Supplementary material}

# Spatio-temporal observation of higher-order modulation instability in a recirculating fiber loop: supplemental document

**CONTINUOUS TRANSITION FROM SEEDED TO SPONTANEOUS MI: EVIDENCES IN THE EVOLUTION OF THE SECOND ORDER MOMENT**

In the letter, we report the observation of a gradual loss of coherence in the spatio-temporal evolution of higher-order MI when the frequency detuning between pump and seed lasers decreases down to 0. This is associated to the emergence of spontaneous MI which progressively dominates the dynamics at the expense of higher-order seeded MI. Qualitative signatures of this interplay appear clearly in the single-shot spatio-temporal diagrams of Fig. 2 of the letter. An interesting quantitative insight into this complex interplay is obtained by looking at some statistical indicator and also, by making use of comparisons with numerical simulations. In particular, we focus on the longitudinal evolution of the second-order moment of the optical power $\kappa_4(z)$:

$$\kappa_4(z) = \frac{\langle P(z,t)^2 \rangle}{\langle P(z,t) \rangle^2} \tag{S1}$$

where $P(z,t) = |\Psi(z,t)|^2$ is the optical power and $\langle \cdot \rangle$ denotes average over time $t$. In Fig. 3 of the letter, we compare the experimental measurements of $\kappa_4(z)$ to numerical simulations of the NLS equation (Equation (1)) for 5 different values of the frequency detuning. Numerical simulations do not include random noise on the initial condition such that the difference between experiments and numerics quantitatively illustrates the influence of spontaneous MI in our experiments.

We have recorded more than 200 single-shot spatio-temporal diagrams by sweeping the frequency detuning $\Delta f$ in the range $\pm 13$ GHz with a regular step of $\sim 125$ MHz.

We have performed the computation of $\kappa_4(z)$ for every recording and the result is presented in Fig. S1(a) as a 2D color plot. Each vertical line of this figure represents the longitudinal evolution of $\kappa_4$ over 600 km of propagation calculated from one spatio-temporal recording at a given $\Delta f$. Bright areas (large $\kappa_4$) traduce the existence of a larger number of high power localised structures. Figure S1 clearly reveals a continuous evolution of $\kappa_4(z)$ as a function of $\Delta f$ which illustrates the interplay between higher-order seeded MI and spontaneous MI as discussed in the letter. Below is a list of noteworthy remarks that can be made:

- The results are almost perfectly symetrical for positive and negative frequency detunings. Discrepancies come from small fluctuations of experimental parameters in the course of the run of acquisitions ($\sim 2$ h);

- For $|\Delta f| > 5$ GHz, well defined bands are observed that translate the quasi-periodic longitudinal dynamics of 1$^{\text{st}}$-order seeded MI;

- For $\sim 2$ GHz $< |\Delta f| < 5$ GHz, a more complex band structure appears. In particular, 3 bands exist within the first 300 km of propagation associated in the spatio-temporal dynamics to the first maximum compression, coherent splitting in two and three respectively;

- For $|\Delta f| < \sim 2$ GHz, the bands vanish into a unique region where $\kappa_4$ is almost uniform when propagation distance is longer than $\sim 150$ km. It is in this range of $\Delta f$ that higher-order seeded MI and spontaneous MI exhibit a competition in the spatio-temporal dynamics. In particular, the longitudinal evolution of $\kappa_4(z)$ for $\Delta f \sim 0$ is reminiscent of the one observed in the case of spontaneous MI [1];

For comparison, the same 2D color plot has been constructed from corresponding numerical simulations of the NLS equation without inclusion of noise in the initial condition, and is presented using the same figure layout in Fig. S2. The agreement is remarkable as can be seen from the details within each bands for $\Delta f > 2\,\text{GHz}$ which are also captured in our experiments. Careful comparison eventually reveals that around $\Delta f = 0$, $\kappa_4(z)$ remains equal to 1 in the simulations because no spectral noise in the initial state triggers spontaneous MI. This confirms that the broad uniform region in Fig. S1 around $\Delta f = 0$ find its origin in the emergence of spontaneous MI in the spatio-temporal dynamics of the system.

**Fig. S1.** (a) Longitudinal evolution of the second-order moment $\kappa_4(z)$ as a function of the frequency detuning $\Delta f$ computed from the experimental recordings. (b-i) $\kappa_4(z)$ curves extracted from (a) for values of $\Delta f$ indicated by the dashed gray lines.

**Fig. S2.** Same as Fig. S1 with numerical simulations of the NLS equation including realistic losses but no spectral noise in the initial conditions.



\end{document}